%% file: LKB.tex
\documentclass[usenatbib]{mn2e}

\usepackage{footmisc}
\usepackage{psfig}
\usepackage{amsmath}
\usepackage{graphics}
\usepackage{lscape}
\usepackage{times}
\begin{document}

\title[Lutz-Kelker bias in pulsar parallaxes]{Lutz-Kelker bias in
  pulsar parallax measurements}

\author[J.~P.~W.~Verbiest, et al.]
       {J.~P.~W.~Verbiest,$^1$\thanks{E-mail:
           Joris.Verbiest@mail.wvu.edu.} D.~R.~Lorimer,$^1$
         M.~A.~McLaughlin$^1$\\ 
         $^1$Department of Physics, West Virginia University, PO Box
         6315 Morgantown, WV 26506-6315, USA}
\maketitle

\begin{abstract}
  \citeauthor{lk73} showed that parallax measurements are
  systematically overestimated because they do not properly account
  for the larger volume of space that is sampled at smaller parallax
  values. We apply their analysis to neutron stars, incorporating the
  bias introduced by the intrinsic radio luminosity function and a
  realistic Galactic population model for neutron stars. We estimate
  the bias for all published neutron star parallax measurements and
  find that measurements with less than $\sim 95\%$ certainty, are
  likely to be significantly biased.  Through inspection of historic
  parallax measurements, we confirm the described effects in optical
  and radio measurements, as well as in distance estimates based on
  interstellar dispersion measures. The potential impact on future
  tests of relativistic gravity through pulsar timing and on
  X-ray--based estimates of neutron star radii is briefly discussed.
\end{abstract}


\section{Introduction}\label{sec:intro}
The past decade has seen an exponential increase in the number of
parallax measurements to radio pulsars. Whereas only 14 such
measurements were known by the end of the year 2000, currently
52 radio pulsars have their distance determined either through VLBI or
pulsar timing measurements (see Tables \ref{tab:curpxs} and
\ref{tab:oldpxs}).

The importance of radio pulsar parallaxes arises from the wide variety
of highly accurate investigations to which pulsars lend
themselves. For example, the combination of an accurate parallax with
the pulsar dispersion measure (DM: the integrated electron density
between the pulsar and Earth) provides an average electron density
measurement along the line of sight, which can be used to construct
Galactic electron density models \citep[see,
e.g.,][]{cl02}. Also, the highly polarised nature of pulsars allows
measurement of the Faraday rotation which - in combination with a
distance - can be used to map out the component of the Galactic
magnetic field parallel to the line of sight \citep{hml+06}. Finally,
in pulsar timing, distances are essential in correcting spindown and
orbital period derivatives for the Shklovskii effect caused by proper
motion \citep{shk70}.  Consequentially, certain pulsar timing tests of
general relativity are dependent on accurate distance measurements, as
described by \citet{dt91} and used in \citet{nss+05} and
\citet{dvtb08}, amongst others.

The large increase in the number of radio pulsar parallaxes and their
unique applications warrant an investigation into potential biases. As
pointed out by \citet{lk73}, in an homogeneous field of stars the
number of stars per unit of distance increases as $D^2$, where $D$ is
distance. This makes it statistically more likely to find an object at
larger distances where more volume is sampled and, hence, more sources
lie. \citet{lk73} showed analytically that this statistical
underestimate of the stellar distance depends on the precision of the
parallax measurement. Specifically, they derived the following
proportionality:
\begin{equation}
  p(\varpi | \varpi_{\rm 0}) \propto \left( \frac{\varpi_{\rm
      0}}{\varpi}\right)^4 \exp \left( -
  \frac{\left(\varpi-\varpi_{\rm 0}\right)^2}{2 \sigma^2}\right),
\end{equation}
where $\varpi_{\rm 0}$ is the measured parallax, $\sigma$ is the
standard deviation of the measurement and $p(\varpi | \varpi_{\rm 0})$
is the probability distribution of the actual parallax, $\varpi$,
given the measurement. \citet{bm98} expanded this analysis by using
the intrinsic luminosity function for the star's spectral class as a
further source of prior information to use in the estimate of a more
accurate parallax value.

Two points of confusion have pervaded the literature on the topic of
Lutz-Kelker bias. First, the effect we describe as Lutz-Kelker bias is
commonly referred to as Malmquist bias in extragalactic astronomy. As
\citet{gf97} point out, the Malmquist bias originally referred to a
positive bias in luminosity of magnitude-limited samples and has only
recently acquired the meaning of the geometric bias described by
\citet{lk73}. \citet{smi03b} stresses a second point of confusion and
difference between the original Malmquist bias and Lutz-Kelker bias,
namely the fact that the bias described by \citet{lk73} and discussed
in the present paper, refer to individual parallax measurements rather
than to the overall average of a sample of objects.

In this paper we revisit the analysis by \citet{bm98} with a
particular focus on radio pulsars. To account for the analytically
complex but realistic pulsar luminosity function and Galactic pulsar
distribution, we describe a Monte-Carlo approach to correct previously
published parallax measurements for the Lutz-Kelker bias. Our basic
analytic derivation and a description of the simulations are given in
\S\ref{sec:theory}. The resulting corrections to published
measurements are discussed in \S\ref{sec:revise}. Our conclusions are
summarised in \S\ref{sec:conc}.

\section{Theory and Simulations}\label{sec:theory}
As an illustration of the bias caused by volumetric and intrinsic
luminosity arguments, it is helpful to consider an analytic derivation
along the lines of those presented by \citet{lk73} and
\citet{bm98}. Such an analysis is presented in \S\ref{ssec:Basis}
below. For any practical purposes, however, a more realistic approach
that incorporates a Galactic distribution model for neutron stars as
well as a more realistic intrinsic luminosity function for radio
pulsars, is required. Because such realistic models are analytically
complex, we will implement them only as part of a Monte-Carlo
simulation, which is described in \S\ref{ssec:sims}.

\subsection{Basic Theory}\label{ssec:Basis}
For a pulsar at distance $D = \varpi^{-1}$, Bayes' theorem gives the
probability that the true parallax is $\varpi$, given the measurement
$\varpi_{\rm 0}$, as
\begin{equation}\label{eq:Bayes}
  p(\varpi | \varpi_{\rm 0}) = \frac{p(\varpi_{\rm 0} | \varpi)
    p(\varpi)} {p(\varpi_{\rm 0})},
\end{equation}
where $p(\varpi_{\rm 0})$ is the prior probability of the measurement
which is taken to be constant without loss of generality. Assuming the
measurement is dominated by Gaussian noise with standard deviation
$\sigma$, we know furthermore that
\begin{equation}\label{eq:Gauss}
  p(\varpi_{\rm 0}|\varpi) = \frac{1}{\sqrt{2
      \pi}\sigma}\exp\left[-\frac{1}{2}\left(\frac{\varpi-\varpi_{\rm
      0}}{\sigma}\right)^2\right].
\end{equation}
The final term in Eq. \ref{eq:Bayes} is the prior probability
distribution of the parallax, $p(\varpi)$, which is composed of two
terms: one defined by the sampled volume and another defined by the
intrinsic pulsar luminosity function. Specifically, assuming a
homogeneous distribution of pulsars in space, we have
\[
  p(D) \propto \rho D^2,
\]
where $\rho$ is the spatial density of pulsars. Translating this
probability to a probability in $\varpi$, we get
\begin{equation}\label{eq:volume}
  p_{\rm D}(\varpi) = \left|\frac{\partial D}{\partial \varpi}\right| p(D)
  \propto \varpi^{-4}.
\end{equation}
Note this is effectively the prior \citet{lk73} determined. We will
henceforth refer to this prior as the ``volumetric'' prior
distribution.

The second contribution to the parallax prior is determined by the
intrinsic radio pulsar luminosity function. Assuming, for simplicity,
the power-law luminosity function
\[
  p(L) \propto L^{\beta},
\]
where $L$ is the intrinsic pulsar luminosity and $\beta$ is the slope
of the luminosity function, we obtain
\begin{equation}\label{eq:lumin}
  p_{\rm L}(\varpi) = \left|\frac{\partial L}{\partial \varpi}\right|
  p(L) \propto \varpi^{-2\beta-3},
\end{equation}
which we will refer to as the ``luminosity'' prior distribution.
Combination of Eqs. \ref{eq:volume} and \ref{eq:lumin} results in the
prior probability distribution of the pulsar parallax
\[
  p(\varpi) = p_{\rm D}(\varpi) p_{\rm L}(\varpi) \propto \varpi^{-2\beta-7}.
\]
Inserting this along with Eq. \ref{eq:Gauss} into Eq.  \ref{eq:Bayes},
we obtain the resulting probability distribution of the true parallax:
\begin{equation}\label{eq:basres}
  p(\varpi | \varpi_{\rm 0}) \propto \varpi^{-2\beta-7} {\rm
    e}^{-\frac{1}{2}\left(\frac{\varpi_{\rm
        0}-\varpi}{\sigma}\right)^2}.
\end{equation}
The (local) maximum of this probability distribution can easily be
determined by requiring
\[
  \frac{\partial p(\varpi |\varpi_{\rm 0})}{\partial \varpi} = 0,
\]
which gives:
\begin{multline*} 
    (-2\beta-7)\varpi^{-2\beta-8}{\rm e}^{-\frac{1}{2}\left(
        \frac{\varpi_{\rm 0}-\varpi}{\sigma}\right)^2}  \\
    + \varpi^{-2\beta-7}\frac{\varpi_{\rm
        0}-\varpi}{\sigma^2}{\rm e}^{-\frac{1}{2}\left(
        \frac{\varpi_{\rm 0}-\varpi}{\sigma}\right)^2}  = 0.
\end{multline*}
Rearranging this expression results in the analytic statement of the
bias:
\begin{equation}\label{eq:bassol}
  \frac{\varpi}{\sigma} = \frac{1}{2}\left[\frac{\varpi_{\rm
        0}}{\sigma} \pm \sqrt{\frac{\varpi_{\rm
          0}^2}{\sigma^2}-4(2\beta+7)}\right].
\end{equation}
This result clearly demonstrates that the bias is a function of the
significance of the parallax measurement and not of the parallax
itself.

Equation \ref{eq:bassol} also shows the effect of different intrinsic
luminosity functions: for steep power laws ($\beta < -3.5$) the
luminosity function dominates the prior and the measured parallax
value is likely to be underestimated. For shallow luminosity functions
($\beta > -3.5$) the reverse is true and the parallax is expected to
be overestimated. Note that in this case the prior in Equation
\ref{eq:basres} goes to infinity as $\varpi$ goes to $0$. A luminosity
function with spectral index $\beta = -3.5$ results in a luminosity
prior that exactly cancels out the volumetric prior, so that no bias
is observed (as can easily be seen in Equation \ref{eq:basres}). For
the observed pulsar population, $\beta \approx -1.7$ \citep{lfl+06},
which would imply measured parallaxes to be larger than their actual
values.

\subsection{Realistic Monte-Carlo Simulation}\label{ssec:sims}
\begin{figure}
  \psfig{angle=0.0,width=8.5cm,figure=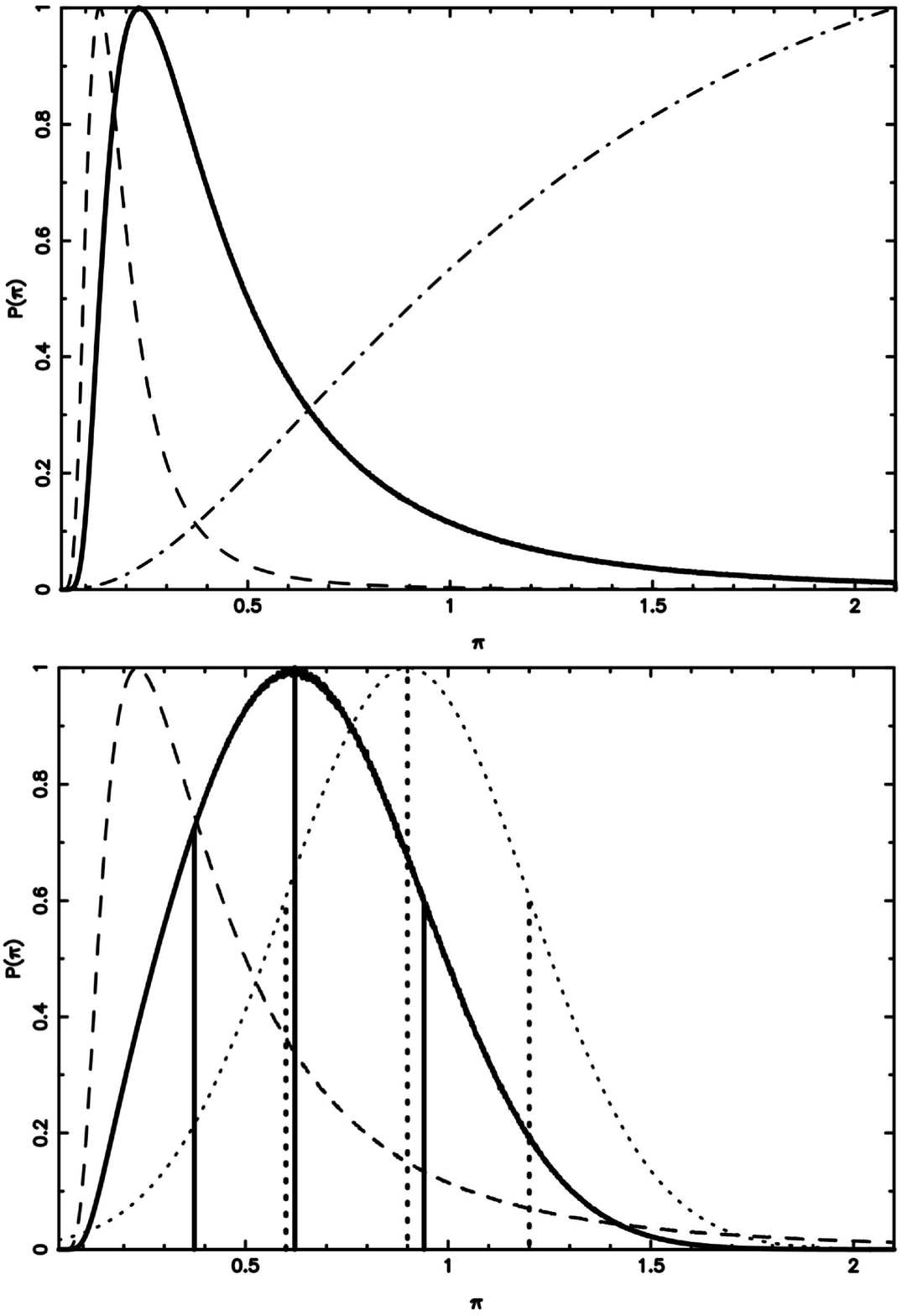}
  \caption{Example output from our Monte-Carlo simulations, applied to
    the PSR J1713+0747 parallax measurement of \citet{cfw94}
    ($\varpi_{\rm 0} = 0.9 \pm 0.3\,$mas). The top figure shows the
    volumetric and luminosity prior distributions (dashed and
    dot-dashed lines respectively) as well as their normalised
    product, the total prior distribution (solid line). The bottom
    figure shows the original measurement (dotted line), the prior
    distribution (dashed line) and their normalised product: the
    bias-corrected probability distribution (solid line). Vertical
    lines show the peak and 1$\,\sigma$ error bars of the measured
    (dotted) and bias-corrected (solid line) probability
    distributions.}
  \label{fig:sample}
\end{figure}

There are two clear improvements to be made to the analysis presented
in the previous section. First, \citet{fk06} have shown that the
intrinsic luminosity function of radio pulsars is not a simple
power-law but more likely log-normal in form. Second, pulsars are not
distributed homogeneously in space, but are mostly confined to the
Galactic disk. The analytic complications implied by these
improvements require that we proceed to evaluate the bias through
Monte-Carlo simulations. Practically, we simulate the Gaussian
measurement uncertainty and the two prior distributions independently,
creating three normalised probability distributions\footnote{Since
  both the scale height and intrinsic luminosity may be argued to
  depend on pulsar age, a more advanced, joint, simulation might be
  called for. Given current uncertainties in pulsar luminosities and
  the luminosity-age relation, however, we exclude this extension from
  the present analysis.}. Multiplication of these distributions
provides the final, bias-corrected, distribution,
$p(\varpi|\varpi_{\rm 0})$ (see for example Figure
\ref{fig:sample}). We will report the peak of this combined
probability distribution as the bias-corrected parallax value,
$\varpi_{\rm Corr}$.

In simulating the volumetric prior, we model a Galaxy of pulsars with
a radius of $15\,$kpc and with the radial density profile derived by
\citet{lfl+06} (their Equation 10, Model C fit):
\[
  \rho(R) = A\left(\frac{R}{\rm R_{\odot}}\right)^B
  \exp\left[-C\frac{R-{\rm R_{\odot}}}{\rm R_{\odot}}\right],
\]
with constants $A = 41\,$kpc$^{-2}$, $B = 1.9$, $C = 5$ and
R$_{\odot}=8.5\,$kpc (i.e. the distance between the Sun and the
Galactic centre). The density distribution above the Galactic plane is
modelled by an exponential distribution $p(z) \propto \exp(-|z|/E)$ where
$E$ is the scale height, taken to be 330\,pc for common pulsars
\citep[in accordance with the findings of][]{lfl+06} and 500\,pc for
millisecond pulsars \citep{lor95,cc97}. To optimise the volumetric
prior, we use the simulated Galaxy of pulsars to determine a
volumetric prior probability density function for sky sectors of
10$^{\circ}$ in Galactic latitude and 15$^{\circ}$ in Galactic
longitude and for the polar regions, defined as having a Galactic
latitude in excess of 85$^{\circ}$.

The prior information of the intrinsic radio pulsar luminosity
function is simulated through random realisations of pulsar
luminosities from the log-normal distribution derived by \citet{fk06},
with mean intrinsic luminosity 0.07\,mJy\,kpc$^2$ (i.e. $\langle \log
L\rangle = -1.1$) and standard deviation $\sigma_{\log L} = 0.9$ in
base-10 logarithms. This luminosity distribution is subsequently
converted into a distribution of parallaxes using the known pulsar
flux, $S$, via: $S \varpi^{-2} = L$. The flux values we used are given
in column 3 of Table \ref{tab:curpxs}. Notice that, unlike the
power-law luminosity function described in Section \ref{ssec:Basis},
the log-normal distribution does have a maximum. This implies that,
depending on the pulsar flux, parallaxes can be biased towards larger
as well as smaller values.

Our software for estimation of the bias is available
on-line\footnote{http://psrpop.phys.wvu.edu/LKbias}, both through a
web interface and for download and off-line execution.

\section{Parallax Measurement Revisions}\label{sec:revise}
We have determined the bias-corrected confidence intervals for all
known VLBI and timing parallax measurements found in the ATNF pulsar
catalogue \citep{mhth05}, on Shami Chatterjee's dedicated
webpage\footnote{Shami Chatterjee's pulsar parallax webpage can be
  found at: http://www.astro.cornell.edu/research/parallax.}  and
through NASA's Astrophysics Data System Bibliographic Services. These
results can be found in Tables \ref{tab:curpxs} and
\ref{tab:oldpxs}. Note that in these tables and throughout the paper,
we quote 1\,$\sigma$ error bars.

In order to evaluate the potential of our bias-correction method, we
compare historic measurements and their corrected values to more
precise, recent measurements in Section \ref{ssec:history}. Next, we
discuss the most significant revisions presented, in Section
\ref{ssec:Significant}. Possible effects on models of the interstellar
medium density and distances derived from these models are discussed
in Section \ref{ssec:ISM}. In Section \ref{ssec:specific} the
potential impact of Lutz-Kelker bias on systems of particular interest
for fundamental tests of gravitational theories is described.
Finally, bias corrections for some non-radio pulsars are considered in
Section \ref{ssec:nonRadio}.

\input{TableA1}
\input{TableA2}

\subsection{Historic Parallax Measurements}\label{ssec:history}
As discussed in \S\ref{ssec:Basis} and shown by Equation
\ref{eq:bassol}, the size of the Lutz-Kelker bias is strongly
dependent on the significance of the input parallax measurement. This
implies that, as increasingly precise parallax measurements are
published, those parallax values will converge towards decreasing
bias. Comparing the historic parallax measurements collated in Table
\ref{tab:oldpxs} to their most precise updated values, we can now
assess if these corrections have been towards the peak of the prior
distribution, or away from it, respectively indicated by ``T'' and
``A'' in column 5 of Table \ref{tab:oldpxs}. (The peak of the prior
probability density function is determined from the pulsar radio flux
and the volumetric probability density function towards the pulsar and
is provided as $\varpi_{\rm Prior}$ in column 5 of Table
\ref{tab:curpxs}.)

This inspection shows that 31 out of 43 revisions were in the
direction predicted by the prior distributions. The chance of this
happening at random is less than $0.5\%$ so this confirms the
described bias effects are present in the data. It does not, however,
imply bias-correction can resolve all these variations. Specifically,
measurements with underestimated uncertainties will be
under-corrected, as is supposedly the case for the PSR J0953+0755
measurement by \citet{gtwr86} \citep[see \S4 in][for a discussion of
this measurement]{bbb+00}. We note that nearly half of all revisions
listed in Table \ref{tab:oldpxs} are significant at the 1\,$\sigma$
level, irrespective of bias-correction. This strongly suggests more
wide-spread underestimation of parallax uncertainties. Also, in cases
where stronger systematic biases are present, application of the
Lutz-Kelker correction will result in equally unreliable results. For
PSR J0437$-$4715 for example, \citet{vbv+08} have demonstrated that
the parallax value published by \citet{vbb+01} differed from later
measurements because of inaccuracies in the Solar System Ephemeris
models used.

\subsection{Significant Parallax Revisions}\label{ssec:Significant}
As follows from Table \ref{tab:curpxs}, out of the 56 pulsars that
currently have parallax measurements from radio or optical data, only
9 have a bias larger than one standard deviation. In the case of PSRs
J0633+1746 and J0720$-$3125 this is mostly due to the absence of
intrinsic luminosity information, as described in Section
\ref{ssec:nonRadio}, while for PSR J0751+1807 the parallax bias is
mostly caused by the low significance of the measurement. For the
remaining 6 pulsars the significance of the bias is caused by the fact
that the luminosity and volumetric prior peak on the same side of the
measured value. This causes the two components to the prior to
reinforce each other, resulting in more significant bias corrections.

The bias correction of PSR J0108$-$1431 is of particular interest,
since this pulsar was once assumed to be the closest known neutron
star \citep{tnj+94}, based on its DM distance of $D_{\rm 1994} =
130\,$pc as derived from the Galactic electron density model of
\citet{tc93}. The updated model of \citet{cl02} increased the distance
to $D_{\rm 2002} = 200\,$pc and a recent VLBI parallax \citep{dtbr09}
further increased this value to $D_{\rm 2009} =
240^{+124}_{-61}\,$pc. Our bias-corrected parallax $\varpi_{\rm Corr}
= 1.6^{+1.6}_{-0.6}$ places the pulsar at a distance of $D_{\rm
  J0108-1431} = 625_{-313}^{+375}\,$pc, strongly suggesting that the
3\,$\sigma$ parallax measurement of \citet{dtbr09} is still
underestimating the distance. 

The bias-corrected parallax of PSR B1541+09 is 1.5\,$\sigma$ larger
than the VLBI value. This pulsar has a Galactic latitude of
46$^{\circ}$, which in combination with the parallax measurement of
$\varpi_{\rm 0} = 0.13 \pm 0.02$ \citep{cbv+09} would place it at a
height of 5.5\,kpc above the Galactic disk. This is a highly unlikely
height given the scale height of 330\,pc for normal pulsars,
suggesting that either the parallax measurement is strongly
underestimated because of bias effects or systematic errors, or the
scale height in our Galactic model is underestimated near the Galactic bulge.

\subsection{ISM Density Models}\label{ssec:ISM}
As mentioned in Section \ref{sec:intro}, models of the Galactic
electron distribution are based on pulsar parallax measurements
\citep{cl02}. Since the estimated biases for most measurements are
small (as follows from Table \ref{tab:curpxs}) and the uncertainties
in the electron density models are relatively large, bias correction
is not expected to significantly affect such models. In future
modelling efforts, however, bias-corrected parallax values should
result in more reliable models.

Conversely, Galactic electron density models are used to estimate
pulsar distances based on the DM. While these distance estimates are
not trigonometric, the same reasoning as presented in Section
\ref{ssec:Basis} holds and DM-derived distances can be expected to be
affected by Lutz-Kelker bias. In order to investigate the presence of
Lutz-Kelker bias in DM distances, we consider all pulsars with
trigonometric parallaxes that were not contained in the most recent
Galactic electron density model and that have measurement
uncertainties lower than 20\% \citep[their Figure 12]{cl03}. The DM
distances and parallax values for these 21 pulsars are compiled in
Table \ref{tab:DM}, assuming a 20\% uncertainty on the DM
distance. Comparison of the DM distances with the more precise
parallax measurements from pulsar timing and VLBI compiled in Table
\ref{tab:curpxs} shows that in 15 of the 21 cases the more precise
measurement lies closer to the peak of the prior distribution
$\varpi_{\rm Prior}$, indicating the presence of Lutz-Kelker
bias. Bias-corrected DM-derived parallaxes are also provided, though
interpretation of these values is complicated by the fact that the
uncertainties of DM distances are rough estimates. This analysis
demonstrates that DM distances are likely to be underestimated, a fact
that will need to be taken into account in research that makes use of
such distances, for example in the determination of pulsar velocities
as in \citet{hllk05}.  \input{Table3}

\subsection{Binary Pulsars}\label{ssec:specific}

One particular application where bias correction may prove crucial in
future research, is the determination of gravitational wave emission
from binary pulsar systems. This emission is measured through orbital
period decay, predicted by general relativity to be \citep{tw82}:
\[
\dot{P}_{\rm b}^{\rm GR} = 
    - \frac{192\pi}{5} 
    \left(
         \frac{2 \pi T_{\odot}}
           {P_{\rm b}}
    \right)^{5/3} 
          \frac{M_{\rm c} M_{\rm p}}{M_{\rm tot}^{1/3}}
\frac{1+\frac{73}{24}e^2+\frac{37}{96}e^4}{\left(1-e^2\right)^{7/2}},
\]
with $\dot{P}_{\rm b}$ the first derivative of the orbital period,
$P_{\rm b}$, $T_{\odot} = 4.925490947 \mu$s, $M_{\rm p}$ the pulsar
mass, $M_{\rm c}$ the companion mass, $M_{\rm tot} = M_{\rm p}+M_{\rm
  c}$ the total system mass and $e$ the orbital eccentricity. Given an
independent measurement of $M_{\rm c}$ and $M_{\rm p}$, a measurement
of $\dot{P}_{\rm b}$ can be used to test gravitational wave emission
theories, as first done by \citet{tw82}. Alternatively, it can be used
to determine the pulsar and companion masses, as in
\citet{nss+05}. Depending on the pulsar's proper motion and position
in the Galaxy, however, certain kinematic terms may contaminate the
$\dot{P}_{\rm b}$ measurement, as first described in detail by
\citet{dt91}. They showed that three extra terms need to be considered
and corrected for: the Shklovskii term based on the pulsar proper
motion and distance \citep{shk70}; the apparent acceleration caused by
the gravitational potential of the Galaxy, dependent on the pulsar's
Galactic latitude and height above the Galactic plane; and the
apparent acceleration caused by differential Galactic rotation, which
is also dependent on the Galactic latitude and therefore the distance
of the pulsar. Since all three non-relativistic contributions to the
measured orbital period derivative $\dot{P}_{\rm b}$ are dependent on
the pulsar distance, accurate estimation of these contaminating
factors and subsequent determination of gravitational wave emission or
pulsar mass estimates, is dependent on an accurate parallax
determination.

Presently, the orbital decay $\dot{P}_{\rm b}$ has only been
determined for 16 binary pulsar systems, of which only four have a
parallax measurement. In the cases of PSRs J0437$-$4715 and
J1909$-$3744 the parallax measurement is precise enough to remain
unaffected by bias effects and in the cases of PSRs J0737$-$3039A/B
and J0751+1807 the uncertainty in the measured $\dot{P}_{\rm b}$ value
is substantially larger than the expected contributions from kinematic
and Galactic effect \citep{ksm+06,nss+05}.

\subsection{Optical Pulsar Parallax Measurements}\label{ssec:nonRadio} 
A total of four neutron stars have had their parallaxes measured
though optical observations with the Hubble Space Telescope (PSRs
J0633+1746, J0720$-$3125, J0835$-$4510 and J1856$-$3754). Of these
four, only the Vela pulsar (PSR J0835$-$4510) has been detected at
radio wavelengths. Since an optical intrinsic luminosity function of
neutron stars has thus far not been determined, we disregard the
luminosity prior for the remaining three optical neutron
stars. Because surveys for radio pulsars are generally biased towards
bright pulsars, most known radio pulsars lie at the bright end of the
intrinsic luminosity function. This implies that the luminosity prior
tends to counterbalance the volumetric prior, which is biased towards
small parallax values (as seen in Figure \ref{fig:sample}). The
exclusion of the luminosity prior for non-radio neutron stars implies
that the effect of the volumetric prior becomes much stronger, which
means higher measurement precision is required before anything
definite can be said about the bias in the neutron star distance. A
clear example of this is PSR J1856$-$3754 for which the 3.5\,$\sigma$
measurement of \citet{kva02} is insufficient to overcome the steep
volumetric prior. For pulsars with luminosity information, a
3.5\,$\sigma$ measurement generally \emph{does} suffice to restrict
the bias within the 68\% confidence interval, as illustrated by the
PSR J0613$-$0200 measurement of \citet{hbo06}. We note that the most
precise parallaxes for PSRs J0633+1746 and J0720$-$3125 are only
3.6\,$\sigma$ and 3.1\,$\sigma$ measurements respectively, which is
insufficient to claim confidence in the inferred distance because of
the poor estimate of the bias.

X-ray luminosities of neutron stars are commonly used to derive
neutron star radii which in turn are used to place constraints on
equations of state for dense nuclear matter. One of the limiting
factors in such analyses are the high distance uncertainties since
distances are mainly derived from mostly inprecise methods such as
interstellar dispersion, supernova remnant associations and
interstellar absorption lines, while only few parallax measurements
are available \citet{plps04}. This lack of precise distances
causes even 2 and 3\,$\sigma$ parallax measurements to
carry significant weight in these analyses. As our work shows, without
prior information on the intrinsic luminosity distribution of these
sources, it cannot be confidently claimed that parallax measurements
with such low significance are accurate enough to be used because they
are likely to underestimate the distance to the neutron star. Based on
equations 56 and 58 as well as Figure 2A of \citet{lp07}, these
underestimated distances can be expected to result in underestimations
of neutron star masses and radii, slightly biasing the analysis to
prefer softer equations of state.

\section{Conclusions}\label{sec:conc}

We have reanalysed bias effects first discussed by \citet{lk73} and
described Monte-Carlo simulations that aim to correct for these
biases. Comparison of historic parallax values to the most recent
measurements confirms that the bias effects are present in
observations, though the correction is complicated by systematic
measurement errors and underestimation of parallax
uncertainties. Correction for the described biases may improve some
pulsar-timing tests of gravitational wave emission and may slightly
influence future density models of the interstellar medium. Finally,
we conclude that optical parallax measurements should be used with
caution since we cannot acurately quantify the possible bias in the
inferred distance estimates.

{\bf \noindent Acknowledgments.}

JPWV, DRL and MAM acknowledge support from a WVEPSCoR research
challenge grant held by the WVU Center for Astrophysics. MAM is an
Alfred P. Sloan Fellow. This work has made extensive use of the ATNF
pulsar catalogue \citep{mhth05} and Shami Chatterjee's webpage of
pulsar parallax measurements:
http://www.astro.cornell.edu/research/parallax. The authors thank the
referee Shami Chatterjee for insightful and constructive
comments. JPWV thanks Adam Deller and George Hobbs for useful
discussions on the subject.

\bibliographystyle{mn2e}
\bibliography{journals,psrrefs,modrefs,crossrefs}

\end{document}

%% file: TableA1.tex
\begin{table*}
  \begin{center}
    \caption{Summary of published parallax values. This table contains
      the most up-to-date parallax measurements of pulsars derived
      from VLBI, pulsar timing and optical astrometry. Given are the
      pulsar name, published parallax value $\varpi_{\rm 0}$, flux at
      1400\,MHz $S_{1400}$, bias-corrected parallax $\varpi_{\rm
        Corr}$, the peak of the prior distribution $\varpi_{\rm
        Prior}$ and any related references. Unless another reference
      is given, the fluxes were compiled from \citet{lylg95} and
      \citet{kxl+98}. Pulsars for which the parallax was determined
      optically are identified by asterisks next to their name. Any
      previous measurements are collated in Table \ref{tab:oldpxs}.}
    \label{tab:curpxs}
    \begin{tabular}{ll|cc|cc|c}
      \hline
      \multicolumn{2}{c}{Pulsar name} & $\varpi_{\rm 0}$       & $S_{1400}$ & $\varpi_{\rm Corr}$    & $\varpi_{\rm Prior}$   & Ref.\\
      Jname        & Bname            & (mas)                  & (mJy)      & (mas)                  & (mas)                  &     \\
      \hline                                                                              
      J0030+0451 &  & 3.3   $\pm$ 0.9        &    0.6     & 1.8$^{+1.0}_{-0.8}$    & 0.5$^{+0.6}_{-0.2}$    & \citet{lkn+06,lzb+00}\\
      J0034$-$0721 & B0031$-$07   & 0.93$^{+0.08}_{-0.07}$ &   11       & 0.93$^{+0.08}_{-0.07}$ & 1.0$^{+1.3}_{-0.3}$    & \citet{cbv+09}\\
      J0108$-$1431 & & 4.2   $\pm$ 1.4        &    1.0     & 1.6$^{+1.5}_{-0.6}$    & 0.9$^{+0.8}_{-0.3}$    & \citet{dtbr09} \\
      J0139+5814 & B0136+57     & 0.37  $\pm$ 0.04       &    4.6     & 0.37 $\pm$ 0.04        & 0.38$^{+0.56}_{-0.13}$ & \citet{cbv+09}\\
      J0332+5434 & B0329+54  & 0.94  $\pm$ 0.11       &  203       & 0.93 $\pm$ 0.11        & 0.61$^{+1.51}_{-0.24}$ & \citet{bbgt02}\\

      J0358+5413 & B0355+54 & 0.91  $\pm$ 0.16       &   23       & 0.87 $\pm$ 0.16        & 0.39$^{+0.73}_{-0.13}$ & \citet{ccv+04}\\
      J0437$-$4715 & & 6.396 $\pm$ 0.054      &  142       & 6.395$^{+0.053}_{-0.055}$& 1.01$^{+1.79}_{-0.37}$ & \citet{dvtb08}\\
      J0452$-$1759 & B0450$-$18   & 0.65$^{+1.4}_{-0.6}$   &    5.3     & 0.69$^{+0.64}_{-0.21}$ & 0.69$^{+0.88}_{-0.22}$ & \citet{cbv+09}\\
      J0454+5543 & B0450+55     & 0.84$^{+0.04}_{-0.05}$ &   13       & 0.84$^{+0.04}_{-0.05}$ & 0.49$^{+0.79}_{-0.17}$ & \citet{cbv+09}\\
      J0538+2817 &  & 0.72$^{+0.12}_{-0.09}$ &    1.9     & 0.69$^{+0.12}_{-0.09}$ & 0.27$^{+0.39}_{-0.08}$ & \citet{cbv+09,lwf+04}\\
      \hline                                                                                              
      J0613$-$0200 &  & 0.80 $\pm$ 0.35        &    1.4     & 0.46$^{+0.33}_{-0.17}$ & 0.31$^{+0.40}_{-0.10}$ & \citet{vbc+09} \\
      J0630$-$2834 & B0628$-$28 & 3.0  $\pm$ 0.4         &   23       & 2.9 $\pm$ 0.4          & 0.65$^{+1.01}_{-0.22}$ & \citet{dtbr09} \\
      J0633+1746$^*$ &  & 6.4  $\pm$ 1.8         &   $-$       & 0.17$^{+0.78}_{-0.01}$ & 0.15$^{+0.20}_{-0.01}$ & \citet{cbmt96}\\
      J0659+1414 & B0656+14 & 3.47 $\pm$ 0.36        &    3.7     & 3.32$^{+0.36}_{-0.37}$ & 0.41$^{+0.57}_{-0.14}$ & \citet{btgg03}\\
      J0720$-$3125$^*$ &  & 2.77 $\pm$ 0.89        &    $-$      & 0.24$^{+1.03}_{-0.08}$ & 0.20$^{+0.25}_{-0.06}$ & \citet{kva07}\\

      J0737$-$3039A & & 0.87 $\pm$ 0.14        &    1.6     & 0.80$^{+0.14}_{-0.15}$ & 0.21$^{+0.29}_{-0.06}$ & \citet{dbt09,bjd+06}\\
      J0751+1807 &  & 1.6  $\pm$ 0.8         &    3.2     & 0.6$^{+0.7}_{-0.2}$    & 0.45$^{+0.59}_{-0.14}$ & \citet{nss+05}\\
      J0814+7429 & B0809+74 & 2.31 $\pm$ 0.04        &   10       & 2.30$^{+0.05}_{-0.04}$ & 0.77$^{+0.97}_{-0.26}$ & \citet{bbgt02}\\
      J0820$-$1350 & B0818$-$13 & 0.51$^{+0.03}_{-0.04}$ &    7       & 0.51$^{+0.03}_{-0.04}$ & 0.40$^{+0.62}_{-0.13}$ & \citet{cbv+09}\\
      J0826+2637 & B0823+26 & 2.8  $\pm$ 0.6         &   10       & 2.5 $\pm$ 0.6          & 0.75$^{+0.99}_{-0.24}$ & \citet{gtwr86}\\
      \hline                                                                                              
      J0835$-$4510$^*$ & B0833$-$45 & 3.5  $\pm$ 0.2         & 1100       & 3.5 $\pm$ 0.2          & 0.83$^{+1.81}_{-0.36}$ & \citet{dlrm03,bf74}\\
      J0922+0638 & B0919+06 & 0.83 $\pm$ 0.13        &    4.2     & 0.83 $\pm$ 0.13        & 0.78$^{+0.89}_{-0.25}$ & \citet{ccl+01}\\
      J0953+0755 & B0950+08 & 3.82  $\pm$ 0.07       &   84       & 3.81 $\pm$ 0.07        & 1.20$^{+1.50}_{-0.42}$ & \citet{bbgt02}\\
      J1012+5307 & & 1.22  $\pm$ 0.26       &    3       & 1.09$^{+0.27}_{-0.23}$ & 0.65$^{+0.76}_{-0.21}$ & \citet{lwj+09}\\
      J1022+1001 & & 1.8   $\pm$ 0.3        &    3       & 1.7 $\pm$ 0.3          & 0.62$^{+0.74}_{-0.19}$ & \citet{vbc+09}\\

      J1024$-$0719 & & 1.9   $\pm$ 0.8        &    0.66    & 0.7$^{+0.7}_{-0.2}$    & 0.45$^{+0.50}_{-0.13}$ & \citet{hbo06}\\
      J1045$-$4509 & & 3.3   $\pm$ 1.9        &    3       & 0.25$^{+0.47}_{-0.08}$ & 0.24$^{+0.35}_{-0.07}$ & \citet{vbc+09}\\
      J1136+1551 & B1133+16 & 2.80  $\pm$ 0.16       &   32       & 2.8$^{+0.15}_{-0.17}$  & 1.29$^{+1.40}_{-0.44}$ & \citet{bbgt02}\\
      J1239+2453 & B1237+25 & 1.16  $\pm$ 0.08       &   10       & 1.17$^{+0.07}_{-0.09}$ & 1.17$^{+1.20}_{-0.39}$ & \citet{bbgt02}\\
      J1300+1240 & B1257+12 & 1.3   $\pm$ 0.4        &    2       & 1.0$^{+0.4}_{-0.3}$    & 0.64$^{+0.72}_{-0.20}$ & \citet{wdk+00}\\
      \hline                                                                                              
      J1456$-$6843 & B1451$-$68 & 2.2   $\pm$ 0.3        &   80       & 2.1 $\pm$ 0.3          & 0.37$^{+0.89}_{-0.13}$ & \citet{bmk+90a,mhm80}\\
      J1509+5531 & B1508+55 & 0.47  $\pm$ 0.03       &    8       & 0.48 $\pm$ 0.03        & 0.82$^{+1.06}_{-0.25}$ & \citet{cbv+09}\\
      J1537+1155 & B1534+12 & 0.98  $\pm$ 0.05       &    0.6     & 0.97 $\pm$ 0.05        & 0.60$^{+0.59}_{-0.18}$ & \citet{sttw02} \\
      J1543+0929 & B1541+09     & 0.13  $\pm$ 0.02       &    5.9     & 0.16 $\pm$ 0.02        & 0.75$^{+0.92}_{-0.24}$ & \citet{cbv+09}\\
      J1559$-$4438 & B1556$-$44 & 0.384 $\pm$ 0.081       &   40       & 0.368$^{+0.081}_{-0.078}$ & 0.22$^{+0.58}_{-0.07}$ & \citet{dtb09,fgl+92}\\

      J1600$-$3053 & & 0.20  $\pm$ 0.15       &    3.2     & 0.21$^{+0.12}_{-0.06}$ & 0.21$^{+0.32}_{-0.06}$ & \citet{vbc+09,jbo+07}\\
      J1643$-$1224 & & 2.2   $\pm$ 0.4        &    4.8     & 1.9 $\pm$ 0.4          & 0.20$^{+0.34}_{-0.05}$ & \citet{vbc+09}\\
      J1713+0747 &  & 0.94  $\pm$ 0.05       &    8       & 0.93 $\pm$ 0.05        & 0.23$^{+0.42}_{-0.07}$ & \citet{vbc+09}\\
      J1744$-$1134 & & 2.4  $\pm$ 0.1         &    3       & 2.4 $\pm$ 0.1          & 0.14$^{+0.18}_{-0.03}$ & \citet{vbc+09}\\
      J1856$-$3754$^*$ & & 6.2  $\pm$ 0.6         &    $-$      & 6.0 $\pm$ 0.6   & 0.17$^{+0.20}_{-0.04}$ & \citet{vk07}\\
      \hline                                                                                              
      J1857+0943 & B1855+09 & 1.1  $\pm$ 0.2         &    5       & 0.9 $\pm$ 0.2          & 0.13$^{+0.16}_{-0.03}$ & \citet{vbc+09}\\
      J1900$-$2600 & B1857$-$26 & 0.5  $\pm$ 0.6         &   13       & 0.18$^{+0.30}_{-0.05}$ & 0.17$^{+0.35}_{-0.04}$ & \citet{fgbc99}\\
      J1909$-$3744 & & 0.79 $\pm$ 0.02        &    3       & 0.79 $\pm$ 0.02        & 0.19$^{+0.30}_{-0.05}$ & \citet{vbc+09,jbv+03}\\
      J1932+1059 & B1929+10  & 2.77 $\pm$ 0.07        &   36       & 2.76 $\pm$ 0.07        & 0.19$^{+0.44}_{-0.06}$ & \citet{ccv+04,hfs+04}\\
      J1935+1616 & B1933+16 & 0.22$^{+0.8}_{-0.12}$  &   42       & 0.20$^{+0.33}_{-0.06}$ & 0.19$^{+0.46}_{-0.06}$ & \citet{cbv+09}\\

      J1939+2134 & B1937+21 & 0.13 $\pm$ 0.07        &   10       & 0.14$^{+0.05}_{-0.03}$ & 0.15$^{+0.25}_{-0.04}$ & \citet{vbc+09}\\
      J2018+2839 & B2016+28 & 1.0 $\pm$ 0.1          &   30       & 1.0 $\pm$ 0.1        & 0.22$^{+0.49}_{-0.07}$ & \citet{bbgt02}\\
      J2022+2854 & B2020+28 & 0.37 $\pm$ 0.12        &   38       & 0.34$^{+0.12}_{-0.10}$ & 0.23$^{+0.53}_{-0.07}$ & \citet{bbgt02,gg74}\\
      J2022+5154 & B2021+51 & 0.50 $\pm$ 0.07        &   27       & 0.50 $\pm$ 0.07        & 0.39$^{+0.74}_{-0.13}$ & \citet{bbgt02}\\
      J2048$-$1616 & B2045$-$16 & 1.05$^{+0.03}_{-0.02}$ &   13       & 1.05$^{+0.03}_{-0.02}$ & 0.56$^{+0.85}_{-0.18}$ & \citet{cbv+09}\\
      \hline                                                                                              
    \end{tabular}
  \end{center}
\end{table*}
\begin{table*}
  \begin{center}
    \contcaption{}
    \begin{tabular}{ll|cc|cc|c}
      \hline
      \multicolumn{2}{c}{Pulsar name}       & $\varpi_{\rm 0}$        & $S_{1400}$  & $\varpi_{\rm Corr}$     & $\varpi_{\rm Prior}$   & Ref.\\
      Jname & Bname      & (mas)                  & (mJy)      & (mas)                  & (mas)                & \\
      \hline
      J2055+3630 & B2053+36     & 0.17 $\pm$ 0.03        &    2.6     & 0.18 $\pm$ 0.03        & 0.27$^{+0.39}_{-0.08}$ & \citet{cbv+09}\\
      J2124$-$3358 & & 3.1  $\pm$ 0.6         &    1.6     & 2.7 $\pm$ 0.6          & 0.42$^{+0.54}_{-0.13}$ & \citet{vbc+09}\\
      J2129$-$5721 & & 1.9  $\pm$ 0.9         &    1.4     & 0.4$^{+0.7}_{-0.2}$    & 0.36$^{+0.45}_{-0.11}$ & \citet{vbc+09}\\
      J2144$-$3933 & & 6.05 $\pm$ 0.56        &    0.8     & 5.81$^{+0.55}_{-0.60}$ & 0.60$^{+0.63}_{-0.18}$ & \citet{dtbr09}\\
      J2145$-$0750 & & 1.6  $\pm$ 0.3         &    8       & 1.5 $\pm$ 0.3         & 0.52$^{+0.72}_{-0.17}$ & \citet{vbc+09}\\
      J2157+4017 & B2154+40     & 0.28 $\pm$ 0.06        &   17       & 0.29$^{+0.06}_{-0.05}$ & 0.40$^{+0.70}_{-0.14}$ & \citet{cbv+09}\\
      J2313+4253 & B2310+42     & 0.93$^{+0.06}_{-0.07}$ &   15       & 0.92$^{+0.06}_{-0.07}$ & 0.56$^{+0.85}_{-0.19}$ & \citet{cbv+09}\\
      \hline
    \end{tabular}
  \end{center}
\end{table*}

%% file: TableA2.tex
\begin{table*}
  \begin{center}
    \caption{Historic parallax measurements of pulsars. Given are the pulsar name, the published parallax value $\varpi_0$, the significance of the measurement 
      $\varpi_{0}/\sigma$, the change of the measurement with respect to the previously published value $\Delta\varpi_{0}$, T/A indicating whether the measurement was corrected 
      Towards or Away from $\varpi_{\rm Prior}$ (listed in Table \ref{tab:curpxs}) when compared to the most precise measurement available, the bias-corrected parallax value
      $\varpi_{\rm Corr}$, its difference with the previous bias-corrected value $\Delta \varpi_{\rm Corr}$ and the relevant publication.}
    \label{tab:oldpxs}
    \begin{tabular}{l|c c c c|c c|c}
      \hline
      Pulsar       & $\varpi_{\rm 0}$       & Significance & $\Delta\varpi_{\rm 0}$ & Change w.r.t. & $\varpi_{\rm Corr}$ & $\Delta\varpi_{\rm Corr}$ & Ref.    \\
      name         & (mas)                  & $\varpi_{\rm 0}/\sigma$& (mas)  & $\varpi_{\rm Prior}$ & (mas)                  & (mas)                    &               \\
      \hline                                                                                                                   
      J0332+5434   & 1.3  $\pm$ 1.7         &   0.8    & --               & T & 0.7$^{+1.1}_{-0.3}$    & --                       & \citet{sla79} \\
      (B0329+54)   & 0.91 $\pm$ 0.11        &   8.3    & $-0.39 \pm 1.70$ & -- & 0.91 $\pm$ 0.11        & 0.21 $\pm$ 1.11           & \citet{bbgt02}\\
      J0437$-$4715 & 5.6  $\pm$ 0.8         &   7.0    & --               & A & 5.3 $\pm$ 0.8          & --                       & \citet{sbm+97}\\
                   & 7.19 $\pm$ 0.14        &  51.4    & 1.59 $\pm$ 0.81  & T & 7.16 $\pm$ 0.14        & 1.86 $\pm$ 0.81          & \citet{vbb+01}\\
                   & 6.3  $\pm$ 0.1         &  63.0    & $-0.89 \pm 0.17$ & A & 6.3 $\pm$ 0.1          & $-0.86 \pm 0.17$         & \citet{hbo06} \\
                   & 6.65 $\pm$ 0.51        &  13.0    & 0.35 $\pm$ 0.52  & T & 6.58$^{+0.48}_{-0.54}$ & 0.28 $\pm$ 0.55          & \citet{vbv+08}\\
                   & 6.396$\pm$ 0.054       & 118.4    & $-0.254 \pm 0.513$ & -- & 6.395$^{+0.053}_{-0.055}$& $-0.185 \pm 0.543$   & \citet{dvtb08}\\
      J0538+2817   & 0.68 $\pm$ 0.15        &   4.5    & --               & A & 0.60$^{+0.15}_{-0.14}$ & --                       & \citet{nrb+07}\\
                   & 0.72$^{+0.12}_{-0.09}$ &   6.0    & 0.04 $\pm$ 0.17  & -- & 0.68$^{+0.12}_{-0.09}$ & 0.08 $\pm$ 0.17          & \citet{cbv+09}\\
      J0613$-$0200 & 2.1  $\pm$ 0.6         &   3.5    & --               & T & 1.1$^{+0.7}_{-0.5}$    & --                       & \citet{hbo06} \\
                   & 0.80 $\pm$ 0.35        &   2.3    & $-1.3 \pm 0.7$   & -- & 0.46$^{+0.3}_{-0.17}$  & $-0.64 \pm 0.58$         & \citet{vbc+09}\\
      J0737$-$3039A& 3    $\pm$ 2           &   1.5     & --              & T & 0.21$^{+0.35}_{-0.06}$ & --                       & \citet{ksm+06}\\
                   & 0.87 $\pm$ 0.14        &   6.2    & $-2.13 \pm 2.00$ & -- & 0.80$^{+0.14}_{-0.15}$ & 0.59 $\pm$ 0.38          & \citet{dbt09} \\
      J0814+7429   & 1.8  $\pm$ 7.1         &   0.3    & --               & A & 0.7$^{+1.0}_{-0.3}$    & --                       & \citet{sla79} \\
      (B0809+74)   & 2.31 $\pm$ 0.04        &  57.8    & 0.51 $\pm$ 7.10  & -- & 2.31 $\pm$ 0.04        & 1.61 $\pm$ 1.00          & \citet{bbgt02}\\
      J0835-4510   & 3.4  $\pm$ 0.7         &   4.9    & --               & A & 3.1 $\pm$ 0.7          & --                       & \citet{cdmb01}\\
      (B0833$-$45) & 3.5  $\pm$ 0.2         &  17.5     & 0.1 $\pm$ 0.7   & -- & 3.5 $\pm$ 0.2          & 0.4 $\pm$ 0.7            & \citet{dlrm03}\\
      J0922+0638   & 0.31 $\pm$ 0.14        &   2.2    & --               & T & 0.43$^{+0.11}_{-0.9}$  & --                       & \citet{fgbc99}\\
      (B0919+06)   & 0.83 $\pm$ 0.13        &   6.4    & 0.52 $\pm$ 0.19  & -- & 0.83$^{+0.12}_{-0.13}$ & 0.4 $\pm$ 0.2            & \citet{ccl+01}\\
      J0953+0755   & 8.3  $\pm$ 8.8         &   0.9    & --               & T & 1.2$^{+2.1}_{-0.4}$    & --                       & \citet{sla79} \\
      (B0950+08)   & 7.9  $\pm$ 0.8         &   9.9    & $-0.4 \pm 8.8$   & T & 7.8$^{+0.7}_{-0.9}$    & 6.6 $\pm$ 2.3            & \citet{gtwr86}\\
                   & 3.6  $\pm$ 0.3         &  12.0    & $-4.3 \pm 0.9$   & A & 3.5 $\pm$ 0.3          & $-4.3 \pm 1.0$           & \citet{bbb+00}\\
                   & 3.82 $\pm$ 0.07        &  54.6    & 0.22 $\pm$ 0.31  & -- & 3.81 $\pm$ 0.07        & 0.31 $\pm$ 0.31          & \citet{bbgt02}\\
      J1022+1001   & 3.3  $\pm$ 0.8         &   4.1    & --               & T & 2.4$^{+0.9}_{-0.8}$    & --                       & \citet{hbo04} \\
                   & 2.5  $\pm$ 0.4         &   6.3    & $-0.8 \pm 0.9$   & T & 2.4$^{+0.4}_{-0.5}$    & 0.0 $\pm$ 1.03           & \citet{hbo06} \\
                   & 1.8  $\pm$ 0.3         &   6.0    & $-0.7 \pm 0.5$   & -- & 1.7 $\pm$ 0.3          & $-0.7 \pm 0.58$          & \citet{vbc+09}\\
      \hline                                                                                                         
      J1239+2453   & $-6.2 \pm$ 5.2         & $-1.2$   & --               & T & 1.0$^{+1.0}_{-0.3}$    & --                       & \citet{sla79} \\
      (B1237+25)   & 1.16 $\pm$ 0.08        &  14.5    & 7.36 $\pm$ 5.20  & -- & 1.17$^{+0.07}_{-0.09}$ & 0.17 $\pm$ 1.00          & \citet{bbgt02}\\
      J1509+5531   & 0.415$\pm$ 0.037       &  11.2    & --               & T & 0.430$^{+0.034}_{-0.037}$& --                     & \citet{cvb+05}\\
      (B1508+55)   & 0.47 $\pm$ 0.03        &  15.7    & 0.055 $\pm$ 0.048& -- & 0.48 $\pm$ 0.03        & 0.050 $\pm$ 0.045        & \citet{cbv+09}\\
      J1537+1155   & 0.91$^{+0.02}_{-0.14}$ &   6.5    & --               & A & 0.89$^{+0.02}_{-0.13}$ & --                       & \citet{sac+98}\\
      (B1534+12)   & 0.98 $\pm$ 0.05        &  19.6    & 0.07 $\pm$ 0.05  & -- & 0.97 $\pm$ 0.05        & 0.08 $\pm$ 0.05          & \citet{sttw02}\\
      J1713+0747   & 0.9  $\pm$ 0.3         &   3.0    & --               & A & 0.6$^{+0.3}_{-0.2}$    & --                       & \citet{cfw94} \\
                   & 0.89 $\pm$ 0.08        &  11.1    & $-0.01 \pm 0.31$ & A & 0.87 $\pm$ 0.08        & 0.27 $\pm$ 0.31          & \citet{sns+05}\\
                   & 1.10 $\pm$ 0.05        &  22.0    & 0.21 $\pm$ 0.09  & T & 1.1 $\pm$ 0.05         & 0.23 $\pm$ 0.09          & \citet{hbo06} \\
                   & 0.95$^{+0.06}_{-0.05}$ &  15.8    & $-0.15 \pm 0.08$ & T & 0.94$^{+0.06}_{-0.05}$ & $-0.16 \pm 0.08$         & \citet{cbv+09}\\
                   & 0.94 $\pm$ 0.05        &  18.8    & $-0.01 \pm 0.07$ & -- & 0.93 $\pm$ 0.05        & $-0.01 \pm 0.07$         & \citet{vbc+09}\\
      J1744$-$1134 & 2.8  $\pm$ 0.3         &   9.3    & --               & T & 2.6 $\pm$ 0.3          & --                       & \citet{tbm+99}\\
                   & 2.1  $\pm$ 0.2         &  10.5    & $-0.7 \pm 0.4$   & A & 2.0 $\pm$ 0.2          & $-0.6 \pm 0.4$           & \citet{hbo06} \\
                   & 2.4  $\pm$ 0.1         &  24.0    & $0.3 \pm 0.2$    & -- & 2.4 $\pm$ 0.1          & 0.4 $\pm$ 0.2            & \citet{vbc+09}\\
      J1856-3754   & 16.5 $\pm$ 2.3         &   7.2    & --               & T & 15.2 $\pm$ 2.4         & --                       & \citet{wal01}\\
                   &  7   $\pm$ 2           &   3.5    & $-9.5 \pm 3.1$   & T & 0.17$^{+0.57}_{-0.04}$   & $-15.0 \pm 2.5$          & \citet{kva02}\\
                   &  8.5 $\pm$ 0.9         &   9.4    & 1.5 $\pm$ 2.2    & T & 7.2 $\pm$ 1.6          & 7.03 $\pm$ 1.70          & \citet{wl02}\\
                   &  6.2 $\pm$ 0.6         &  10.3    & $-0.8 \pm 1.1$   & -- & 6.0 $\pm$ 0.6          & 5.83 $\pm$ 1.71          & \citet{vk07}\\
      J1857+0943   & 1.2  $\pm$ 0.5         &   2.4    & --               & T & 0.14$^{+0.36}_{-0.04}$ & --                       & \citet{rt91}  \\
      (B1855+09)   & 1.1  $\pm$ 0.3         &   3.7    & $-0.1 \pm 0.6$   & A & 0.7 $\pm$ 0.4          & 0.56 $\pm$ 0.54          & \citet{ktr94} \\
                   & 1.1  $\pm$ 0.2         &   5.5    & 0.0 $\pm$ 0.4    & -- & 0.9 $\pm$ 0.2          & 0.2 $\pm$ 0.5            & \citet{vbc+09}\\
      J1909$-$3744 & 0.88 $\pm$ 0.03        &  29.3    & --               & T & 0.88 $\pm$ 0.03        & --                       & \citet{jhb+05}\\
                   & 0.88 $\pm$ 0.02        &  44.0    & 0.0 $\pm$ 0.04   & T & 0.88 $\pm$ 0.02        & 0.0 $\pm$ 0.04           & \citet{hbo06} \\
                   & 0.79 $\pm$ 0.02        &  39.5    & $-0.09 \pm 0.03$ & -- & 0.79 $\pm$ 0.02        & $-0.09 \pm 0.03$         & \citet{vbc+09}\\
      J1932+1059   & $21.5  \pm$ 8.0        &   2.7    & --               & T & 0.19$^{+0.58}_{-0.06}$ & --                       & \citet{sla79} \\
      (B1929+10)   & $ 5.0  \pm$ 1.5        &   3.3    & $-16.5 \pm 8.1$  & T & 0.22$^{+2.8}_{-0.07}$  & 0.03 $\pm$ 0.58          & \citet{cam95} \\
                   & $ 3.02 \pm$ 0.09       &  33.6    & $-1.98 \pm 1.50$ & T & 3.01 $\pm$ 0.09        & 2.79 $\pm$ 2.80          & \citet{bbgt02}\\
                   & $ 2.77 \pm$ 0.07       &  39.6    & $-0.25 \pm 0.11$ & -- & 2.76 $\pm$ 0.07        & $-0.25 \pm 0.11$         & \citet{ccv+04}\\
      \hline
    \end{tabular}
  \end{center}
\end{table*}
\begin{table*}
  \begin{center}
    \contcaption{}
    \begin{tabular}{l|c c c c|c c|c}
      \hline
      Pulsar       & $\varpi_{\rm 0}$        & Significance & $\Delta\varpi_{\rm 0}$ & Change w.r.t. & $\varpi_{\rm Corr}$  & $\Delta\varpi_{\rm Corr}$ & Ref.          \\
      name         & (mas)                  & $\varpi_{\rm 0}/\sigma$& (mas)      & $\varpi_{\rm Prior}$    &(mas)                   & (mas)                    &               \\
      \hline                                                                                             
      J1939+2134   & $ 0.12 \pm 0.08$       &   1.5   & --               & T  &0.14$^{+0.06}_{-0.04}$ & --                       & \citet{ktr94} \\
      (B1937+21)   & $ 0.13 \pm 0.065$      &   2.0   & 0.01 $\pm$ 0.10  & --  &0.14$^{+0.05}_{-0.03}$ & 0.0 $\pm$ 0.08           & \citet{vbc+09}\\
      J2018+2839   & $ 0.9  \pm$ 3.6        &   0.3   & --               & A  &0.22$^{+0.48}_{-0.07}$ & --                       & \citet{sla79} \\
      (B2016+28)   & $ 1.03 \pm$ 0.10       &  10.3   & 0.13 $\pm$ 3.60  & --  &1.00 $\pm$ 0.10        & 0.78 $\pm$ 0.49          & \citet{bbgt02}\\
      J2022+2854   & $-1.3  \pm$ 4.6        & $-0.3$  & --               & T  &0.23$^{+0.50}_{-0.07}$ & --                       & \citet{sla79} \\
      (B2020+28)   & $ 0.37 \pm$ 0.12       &   3.1   & 1.67 $\pm$ 4.60  & --  &0.34$^{+0.12}_{-0.10}$ & 0.11 $\pm$ 0.51          & \citet{bbgt02}\\
      J2022+5154   & $ 1.8  \pm$ 4.9        &   0.4   & --               & T  &0.4$^{+0.8}_{-0.1}$    & --                       & \citet{sla79} \\
      (B2021+51)   & $ 0.95 \pm$ 0.37       &   2.6   & $-0.85 \pm 4.91$ & T  &0.71$^{+0.37}_{-0.26}$ & 0.31 $\pm$ 0.84          & \citet{cbs+96}\\
                   & $ 0.50 \pm$ 0.07       &   7.1   & $-0.45 \pm 0.38$ & --  &0.50 $\pm$ 0.07        & $-0.21 \pm 0.27$         & \citet{bbgt02}\\
      J2048$-$1616 & $ 1.71 \pm$ 0.91       &   1.9   & --               & T  &0.73$^{+0.82}_{-0.26}$ & --                       & \citet{dtbr09} \\
      (B2045$-$16) & $1.05^{+0.03}_{-0.02}$ &  35.0   & $-0.66 \pm 0.91$ & --  &1.05$^{+0.03}_{-0.02}$ & 0.32 $\pm$ 0.82          & \citet{cbv+09}\\
      J2124$-$3358 & $4 \pm 2$              &   2.0   & --               & T  &0.5$^{+0.8}_{-0.2}$    & --                       & \citet{hbo06} \\
                   & $3.1 \pm 0.55$         &   5.6   & $-0.9 \pm 2.1$   & --  &2.67$^{+0.56}_{-0.61}$ & 2.17 $\pm$ 1.01          & \citet{vbc+09}\\
      J2145$-$0750 & $2.0 \pm 0.6$          &   3.3   & --               & T  &1.4$^{+0.6}_{-0.5}$    & --                       & \citet{lkd+04}\\
                   & $1.6 \pm 0.25$         &   6.4   & $-0.4 \pm 0.7$   & --  &1.5 $\pm$ 0.3          & 0.1 $\pm 0.7$            & \citet{vbc+09}\\
      \hline
    \end{tabular}
  \end{center}
\end{table*}

%% file: Table3.tex
\begin{table*}
  \begin{center}
    \caption{Lutz-Kelker corrections for DM-derived distance
      estimates. Given are the pulsar name, the DM-derived distance
      according to the NE2001 model for Galactic electron density
      $D_{\rm DM}$ \citep{cl01}, the parallax $\varpi_{\rm DM}$
      derived from $D_{\rm}$, assuming a 20\% uncertainty in the
      DM-distance and the Lutz-Kelker corrected parallax $\varpi_{\rm
        DM,Corr}$. This table only contains distances for pulsars that
      have accurate (relative errors less than 20\%) independent
      distances (from timing or VLBI), which were obtained after the
      NE2001 model was created.}
    \label{tab:DM}
    \begin{tabular}{l|ccc}
      \hline
      Pulsar        & $D_{\rm DM}$ & $\varpi_{\rm DM}$ & $\varpi_{\rm DM,Corr}$ \\
      name          & (kpc)       & (mas)            & (mas)                \\
      \hline        
      J0034$-$0721  & 0.41        & 2.4(5)           & 2.2 $\pm$ 0.5       \\
      J0139+5814    & 2.88        & 0.35(7)          & 0.35$^{+0.07}_{-0.06}$\\
                                                     
      J0437$-$4715  & 0.14        & 7(1)             & 6.3$^{+1.4}_{-1.6}$  \\
      J0454+5543    & 0.67        & 1.5(3)           & 1.4 $\pm$ 0.3      \\
      J0538+2817    & 1.22        & 0.8(2)           & 0.74$^{+0.16}_{-0.17}$\\
      J0737$-$3039A & 0.52        & 1.9(4)           & 1.6 $\pm$ 0.4       \\
      J0820$-$1350  & 1.99        & 0.5(1)           & 0.49$^{+0.1}_{-0.09}$ \\
      J0835$-$4510  & 0.24        & 4.2(8)           & 3.9$^{+0.8}_{-0.9}$   \\
      J1012+5307    & 0.41        & 2.4(5)           & 2.1 $\pm$ 0.5       \\
      J1022+1001    & 0.45        & 2.2(4)           & 1.89 $\pm$ 0.45     \\
      J1509+5531    & 0.99        & 1.0(2)           & 1.0 $\pm$ 0.2       \\
      J1543+0929    & 3.49        & 0.29(6)          & 0.33 $\pm$0.05      \\
      J1713+0747    & 0.89        & 1.1(2)           & 1.0 $\pm$ 0.2       \\
      J1744$-$1134  & 0.41        & 2.4(5)           & 1.9$^{+0.5}_{-0.6}$   \\
      J1909$-$3744  & 0.46        & 2.2(4)           & 1.8 $\pm$ 0.5       \\
      J1932+1059    & 0.34        & 2.9(6)           & 2.5 $\pm$ 0.6       \\
      J2048$-$1616  & 0.56        & 1.8(4)           & 1.6 $\pm$ 0.4       \\
      J2055+3630    & 4.62        & 0.22(4)          & 0.22 $\pm$ 0.04     \\
                                                     
      J2124$-$3358  & 0.27        & 3.7(7)           & 3.2$^{+0.7}_{-0.9}$   \\
      J2145$-$0750  & 0.57        & 1.8(4)           & 1.6 $\pm$ 0.4       \\
      J2313+4253    & 1.25        & 0.8(2)           & 0.78 $\pm$ 0.15     \\
      \hline
    \end{tabular}
  \end{center}
\end{table*}